\documentclass{cup-hpl}
\usepackage[none]{hyphenat}
\usepackage{lipsum}

\usepackage{graphicx}
\usepackage{dcolumn}
\usepackage{color}
\usepackage{txfonts}
\usepackage{microtype}
\usepackage{siunitx}
\usepackage{url}
\usepackage{epstopdf}
\usepackage{indentfirst}
\usepackage[colorlinks, linkcolor=black,anchorcolor=black,citecolor=black,filecolor=black,urlcolor=black,bookmarksopen=true]{hyperref}
\usepackage{textcomp}
\usepackage{mathrsfs}
\usepackage{braket}
\usepackage[caption=false,font=normalsize,labelfont=sf,textfont=sf]{subfig}
\usepackage{xcolor} 
\usepackage{amsmath}
\usepackage{mathalpha}
\usepackage{footmisc}

\usepackage{newtxtext}
\let\oldthanks\thanks
\renewcommand{\thanks}[1]{%
  \oldthanks{\fontfamily{ptm}\selectfont#1}%
}

\begin{document}

\newtheorem{theorem}{Theorem}

\shorttitle{Generation of Polarization-Tunable Hybrid Cylindrical Vector $\gamma$ Rays}
\shortauthor{S. M. Liu et al.}
\title{Generation of Polarization-Tunable Hybrid Cylindrical Vector $\gamma$ Rays from Rotating Electron Beams}
\author[]{Si-Man Liu$^{1, }$\thanks{These authors contributed equally to this work.}\ }


\author[]{Yue Cao$^{1, *}$}
\author[]{Kun Xue\textsuperscript{2, }\thanks{xuekun@xjtu.edu.cn}\ }
\author[1]  {Li-Xiang Hu}
\author[1]  {Xin-Yu Liu}
\author[1]  {Xin-Yan Li}
\author[1]  {Chao-Zhi Li}
\author[1]  {Xin-Rong Xu}
\author[1]  {Ke Liu}
\author[1]  {Wei-Quan Wang}
\author[1]  {De-Bin Zou}
\author[1]  {Yan Yin}
\author[2, 3]  {Jian-Xing Li}
\author[]{Tong-Pu Yu$^{1,}$\thanks{tongpu@nudt.edu.cn}}

\address[1]{College of Science, National University of Defense Technology, Changsha 410073, China}
\address[2]{Ministry of Education Key Laboratory for Nonequilibrium Synthesis and Modulation of Condensed Matter, State Key Laboratory of Electrical Insulation and Power Equipment, Shaanxi Province Key Laboratory of Quantum Information and Quantum Optoelectronic Devices, School of Physics, Xi'an Jiaotong University, Xi'an 710049, China}
\address[3]{Department of Nuclear Physics, China Institute of Atomic Energy, P.O. Box 275(7), Beijing 102413, China}
\begin{abstract}

Cylindrical vector (CV) $\gamma$ rays can introduce spatially structured polarization as a new degree of freedom for fundamental research and practical applications. However, their generation and control remain largely unexplored. Here, we put forward a novel method to generate CV $\gamma$ rays with tunable hybrid polarization via a rotating electron beam interacting with a solid foil. In this process, the beam generates a coherent transition radiation field and subsequently emits $\gamma$ rays through nonlinear Compton scattering. By manipulating the initial azimuthal momentum of the beam, the polarization angle of $\gamma$ rays relative to the transverse momentum can be controlled, yielding tunable hybrid CV polarization states. Three-dimensional spin-resolved particle-in-cell simulations demonstrate continuous tuning of the polarization angle across $(-90^\circ, 90^\circ)$ with a high polarization degree exceeding 60\%. Our work contributes to the development of structured $\gamma$ rays, potentially opening new avenues in high-energy physics, nuclear science, and laboratory astrophysics.

\end{abstract}


\maketitle

\section{Introduction}
The polarization of $\gamma$ rays plays a crucial role in exploring diverse fields, including high-energy physics~\cite{moortgat2008polarized, qin2022ultrafast, bragin2017high, lv2025generation}, nuclear physics~\cite{Uggerhoj2005The, mueller2014novel, fagg1959polarization}, materials science~\cite{Vetter2018Gamma, shimazoe2022imaging}, and laboratory astrophysics~\cite{Laurent2011Polarized, yu2024bright, boehm2017circular}. Specifically, circularly polarized $\gamma$ rays serve as powerful probes of parity violation~\cite{moeini2013parity}, elastic photon-photon scattering~\cite{Micieli2016Compton}, and photoproduction of mesons~\cite{Akbar2017Measurement}. Meanwhile, linearly polarized $\gamma$ rays prove essential in nuclear resonance fluorescence imaging~\cite{bertozzi2005nuclear}, probing linear Breit-Wheeler pair production~\cite{zhao2022signatures, zhao2023angle}, and exploring vacuum birefringence~\cite{bragin2017high, dai2024fermionic, lv2025generation}. 
Beyond these homogeneous polarization states, beams with spatially varying polarization, as a class of structured beams, have attracted considerable attention in optics ~\cite{Rubinsztein2017Roadmap, forbes2021structured, Marco2025Trends}. A particular class is the cylindrical vector (CV) beams~\cite{zhan2009cylindrical, chen2018vectorial}, which include beams with radial, azimuthal, or hybrid polarization (a linear superposition of radial and azimuthal polarization). These distinct polarization states have enabled various applications, including high-resolution imaging and focusing with radially polarized beams~\cite{Liu2022Super, kozawa2018superresolution, dorn2003sharper}, trapping and manipulation of particles with azimuthally polarized beams~\cite{Yang2021Optical, Skelton2013Trapping, Huang2012Optical, Donato2012Optical, Moradi2019Efficient}, and high-capacity optical communications with hybrid polarization states~\cite{liu2025efficient}. Extending CV polarization to the $\gamma$-ray regime would introduce spatial distribution of polarization as a new degree of freedom for $\gamma$-ray applications. Furthermore, the ability to actively control this polarization topology, for instance, by continuously tuning it from radial to azimuthal states, would further unlock unprecedented opportunities in broad fields, such as probing or manipulating nuclear and subnuclear structures~\cite{Iliadis2021Linear, Zilges22PPNP, Thiel2012Well, Gottschall2014First}, and investigating photon emission with radial and azimuthal polarization in laboratory astrophysics~\cite{Prokhorov2024Evidence}.

Despite the potential highlighted above, the tunable generation of such CV $\gamma$ rays remains a significant challenge. In the optical domain, components such as spatial light modulators~\cite{garcia2020efficient, maluenda2013reconfigurable} and Q-plates~\cite{naidoo2016controlled, sanchez2015performance} enable flexible polarization manipulation of CV beams, transforming homogeneously polarized beams into CV beams with continuously tunable polarization. However, these optical components that rely on properties such as birefringence and dichroism are not readily applicable to $\gamma$ rays, as the short wavelengths present challenges for their modulation capabilities~\cite{chen2018vectorial, zhao2025research}.

On the other hand, existing schemes for producing and manipulating polarized $\gamma$ rays primarily focus on circularly or linearly polarized $\gamma$ rays, lacking the capability to generate spatially varying polarization states. For instance, in linear Compton scattering, the polarization of the driving laser can determine the circularly or linearly polarization of $\gamma$ rays~\cite{baier1973radiation, ritus1985quantum, Omori2006efficient, Alexander2008Observation, Petrillo2015Polarization, an2018high, wang2024manipulation, Howell2021International}. In incoherent bremsstrahlung, circularly polarized $\gamma$ rays can be generated using longitudinally spin-polarized electrons interacting with metal targets~\cite{olsen1959photon, Giulietti2008Intense, albert2016applications}, while linearly polarized $\gamma$ rays require electron beam-crystal interactions through coherent bremsstrahlung~\cite{Kuraev2010Bremsstrahlung,ter1972high, Lohmann1994Linearly}. The advent of high-intensity lasers has enabled $\gamma$-ray generation via the nonlinear Compton scattering process, yielding brilliant, high-energy polarized $\gamma$ rays. In this nonlinear regime, the polarization of $\gamma$ rays can be controlled through electron polarization and laser polarization, with longitudinally spin-polarized electrons generating circular polarization~\cite{li2020polarized} and transversely polarized electrons~\cite{li2020polarized} or laser polarization~\cite{Wan2020High, Xue2020Generation} producing linear polarization. However, the generation and control of $\gamma$ rays with spatially varying polarization remains largely unexplored. Our recent work~\cite{cao2025generating} took a first step towards this goal by presenting a potential approach to generating CV $\gamma$ rays. Yet, the continuous tuning of hybrid CV polarization states remains a significant challenge. 

This challenge may be addressed by introducing a new property to the seed electron beam. Recent studies have shown that accelerating electrons with lasers carrying orbital angular momentum, such as Laguerre-Gaussian laser pulses~\cite{Hu2024Rotating, Shi2024Advances}, can produce electron beams with azimuthal momentum. Alternative approaches, such as multiple Gaussian laser pulses interacting with plasmas~\cite{Shi2023Efficient}, or a Gaussian laser pulse interacting with a beam-splitting array~\cite{Zhang2025Plasma}, can also generate such rotating beams, improving experimental feasibility. These electron beams with significant azimuthal momentum offer numerous potential applications, including wakefield acceleration~\cite{Neeraj2015Positron, zhang2016acceleration} and attosecond bunch generation~\cite{Hu2024Rotating}. However, the direct measurement of beam azimuthal momentum is difficult in experiment and has received limited attention~\cite{thaury2013angular}.

\begin{figure}
\centering
\includegraphics[width=1\linewidth]{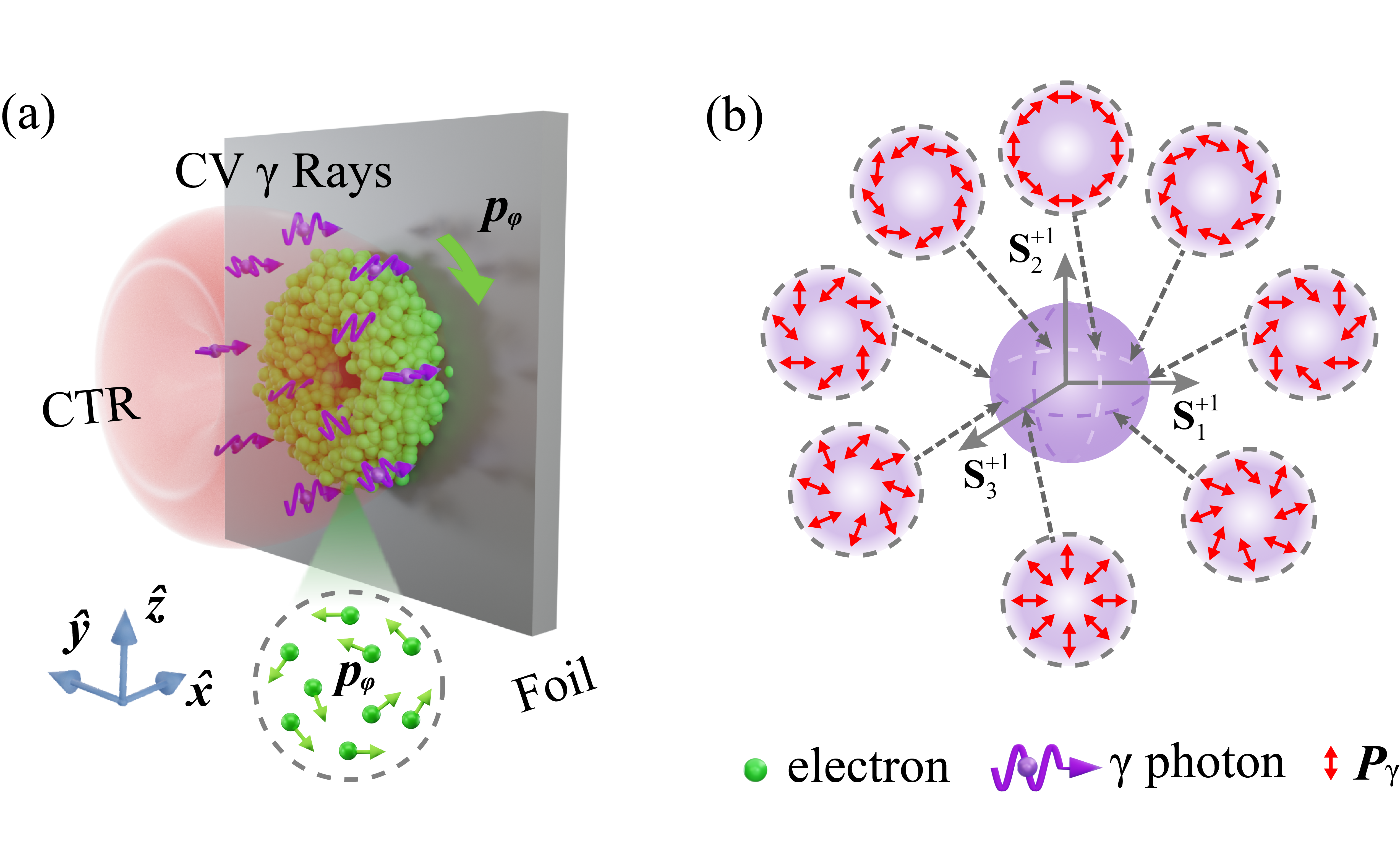}
\caption{Schematic diagram for hybrid cylindrical vector (CV) $\gamma$-ray emission through the interaction of a rotating electron beam and a foil. (a) A relativistic electron beam carrying azimuthal momentum $p_{\varphi}$ propagates in the $+\hat{x}$ direction and traverses a foil, resulting in coherent transition radiation (CTR) and producing polarized $\gamma$ rays through the nonlinear Compton scattering process. The green arrows indicate the azimuthal momentum $p_{\varphi}$ of the beam. (b) Higher-order Poincaré sphere for topological charge $l=1$, where all states on the surface are CV polarized. Points on the equator represent the hybrid mode polarized $\gamma$ rays that can be generated in this scheme. Here, $\mathbf{S}_{3}^{+1} = \pm 1$ represent radial and azimuthal polarization, respectively, while $\mathbf{S}_{1}^{+1} = \pm 1$ correspond to polarization directions at $45^\circ$ and $135^\circ$ with respect to the radial direction, respectively.}
	\label{fig1}
\end{figure}

In this paper, we put forward a novel scheme for the generation of polarization-tunable hybrid CV $\gamma$ rays via the interaction of a rotating electron beam with a solid foil, where the polarization of the emitted photons can be tuned efficiently by the initial azimuthal momentum of the beam. As shown in Figure~\ref{fig1}(a), when the seed electron beam carrying azimuthal momentum interacts with the foil, it generates a coherent transition radiation (CTR) field, which in turn leads to photon emission through the nonlinear Compton scattering process. Manipulating the initial azimuthal momentum of the electrons can control the angle between the polarization direction and transverse momentum of the emitted photons, thereby resulting in tunable hybrid CV polarization states, as shown in Figure~\ref{fig1}(b). Moreover, the relationship between the electron azimuthal momentum and $\gamma$-ray polarization state could also serve as a potential tool for detecting electron azimuthal momentum in laboratory. Three-dimensional spin-resolved quantum electrodynamics (QED) particle-in-cell (PIC) simulations demonstrate the generation of high-energy $\gamma$ rays with a high polarization degree exceeding 60\%. Our work contributes to the active manipulation of the polarization topology of $\gamma$ rays, and could offer a promising tool for investigations in diverse areas, such as high-energy physics~\cite{dai2024fermionic}, nuclear science~\cite{Iliadis2021Linear, Zilges22PPNP, Thiel2012Well, Gottschall2014First}, and laboratory astrophysics~\cite{Prokhorov2024Evidence}.

\section{Simulation Methods and Setup}

We perform simulations using the spin-resolved QED PIC code SLIPs~\cite{wan2023simulations}, which is coupled to a Monte Carlo algorithm for QED processes based on the local constant field approximation~\cite{ritus1985quantum, ilderton2019note,piazza2019,  di2018implementing}. As one of the primary QED processes during the beam-target interaction, the nonlinear Compton scattering process is characterized by the nonlinear QED parameter $\chi_e\equiv|e|\hbar/(m_e^3c^4)\sqrt{-(F_{\mu\nu}p^{\nu})^2}$~\cite{ritus1970radiative, ritus1972radiative, koga2005nonlinear}. Here, $e$ and $m_e$ are the charge and mass of the electron, respectively, $c$ is the speed of light in vacuum, $\hbar$ is the reduced Planck constant, $p^\nu$ is the four-momentum of the electron, and $F_{\mu\nu}$ is the electromagnetic field tensor. When a $\gamma$ photon is emitted, its polarization can be described by the Stokes parameters ($\xi_1,\ \xi_2,\ \xi_3$), which are defined in an instantaneous frame ($\hat{\mathbf{k}}_\gamma$, $\hat{\mathbf{e}}_1$, $\hat{\mathbf{e}}_2$)~\cite{mcmaster1961matrix}, with $\hat{\mathbf{e}}_1=\hat{\mathbf{a}}-\hat{\mathbf{v}}(\hat{\mathbf{v}}\cdot\hat{\mathbf{a}})$ and $\hat{\mathbf{e}}_2=\hat{\mathbf{v}}\times\hat{\mathbf{a}}$. Here, $\hat{\mathbf{a}}$ and $\hat{\mathbf{v}}$ are the unit vectors of the acceleration and velocity of the electron, respectively, and $\hat{\mathbf{k}}_\gamma$ is the unit vector in the direction of photon propagation. The direction of $\hat{\mathbf{k}}_\gamma$ is considered to be aligned with $\mathbf{v}$, since the emission angle is $\sim$ $1/\gamma_e$ $\ll$ 1, where $\gamma_e$ is the Lorentz factor of the electron.  Specifically, $\xi_3$ corresponds to degree of linear polarization along $\hat{\mathbf{e}}_1$ versus $\hat{\mathbf{e}}_2$; $\xi_1$ represents linear polarization at an angle of $\pm45^\circ$ to the $\hat{\mathbf{e}}_1$ axis; and $\xi_2$ denotes degree of circular polarization~\cite{li2019ultrarelativistic}. To detect the mean polarization of photons propagating in a given direction, their Stokes parameters must be transformed to a common observation frame ($\hat{\mathbf{k}}_\gamma$, $\hat{\mathbf{o}}_1$, $\hat{\mathbf{o}}_2$) and then averaged.

The average Stokes parameters of a photon emitted by an electron with initial spin $\mathbf{S}_i$ are given by~\cite{cao2025generating}:
\begin{subequations}
	\label{eq:xi_bar}
	\begin{align}
		\overline{\xi}_{1} &= \frac{{u}/{(1-u)} K_{\frac{1}{3}}(\rho) (\mathbf{S}_i \cdot \hat{\mathbf{a}})}{w - u{ K}_{\frac{1}{3}}(\rho)\hat{{\mathbf b}} \cdot \mathbf{S}_i}, \\
		\overline{\xi}_{2} &= \frac{-\left[u{{\rm Int} K}_{\frac{1}{3}}(\rho)-{(2u-u^2)}/{(1-u)}{ K}_{\frac{2}{3}}(\rho)\right](\mathbf{S}_i\cdot\hat{{\mathbf v}})}{w - u{ K}_{\frac{1}{3}}(\rho)\hat{{\mathbf b}} \cdot \mathbf{S}_i}, \\
		\overline{\xi}_{3} &= \frac{K_{\frac{2}{3}}(\rho) - {u}{(1-u)} K_{\frac{1}{3}}(\rho) (\mathbf{S}_i \cdot \hat{\mathbf{b}})}{w  - u{ K}_{\frac{1}{3}}(\rho)\hat{{\mathbf b}} \cdot \mathbf{S}_i},
	\end{align}
\end{subequations}
where $w = -{{\rm Int} K}_{\frac{1}{3}}(\rho)+\frac{u^2-2u+2}{1-u}{ K}_{\frac{2}{3}}(\rho)$, $\hat{\mathbf{b}} = \hat{\mathbf{v}}\times\hat{\mathbf{a}}/|\hat{\mathbf{v}}\times\hat{\mathbf{a}}|$, $\rho=2u/\left[(1-u)3\chi_e\right]$, $u=\omega_\gamma/\varepsilon_i$ is the ratio of the photon energy $\omega_\gamma$ to the initial electron energy $\varepsilon_i$, and ${{\rm Int} K}_{\frac{1}{3}}(\rho)\equiv \int_{\rho}^{\infty} {\rm d}z { K}_{\frac{1}{3}}(z)$, with ${ K}_n$ the $n$-order modified Bessel function of the second kind. The degree of polarization for each propagation direction $\hat{\mathbf{k}}_\gamma$ is given by $P_\gamma \equiv \sqrt{\overline{\xi_1}^2 + \overline{\xi_3}^2}$. 

\begin{figure*}	[t]	 
	\setlength{\abovecaptionskip}{-0.2cm}
	\setlength{\belowcaptionskip}{-0.3cm}
	\centering
	\includegraphics[width=0.7\linewidth]{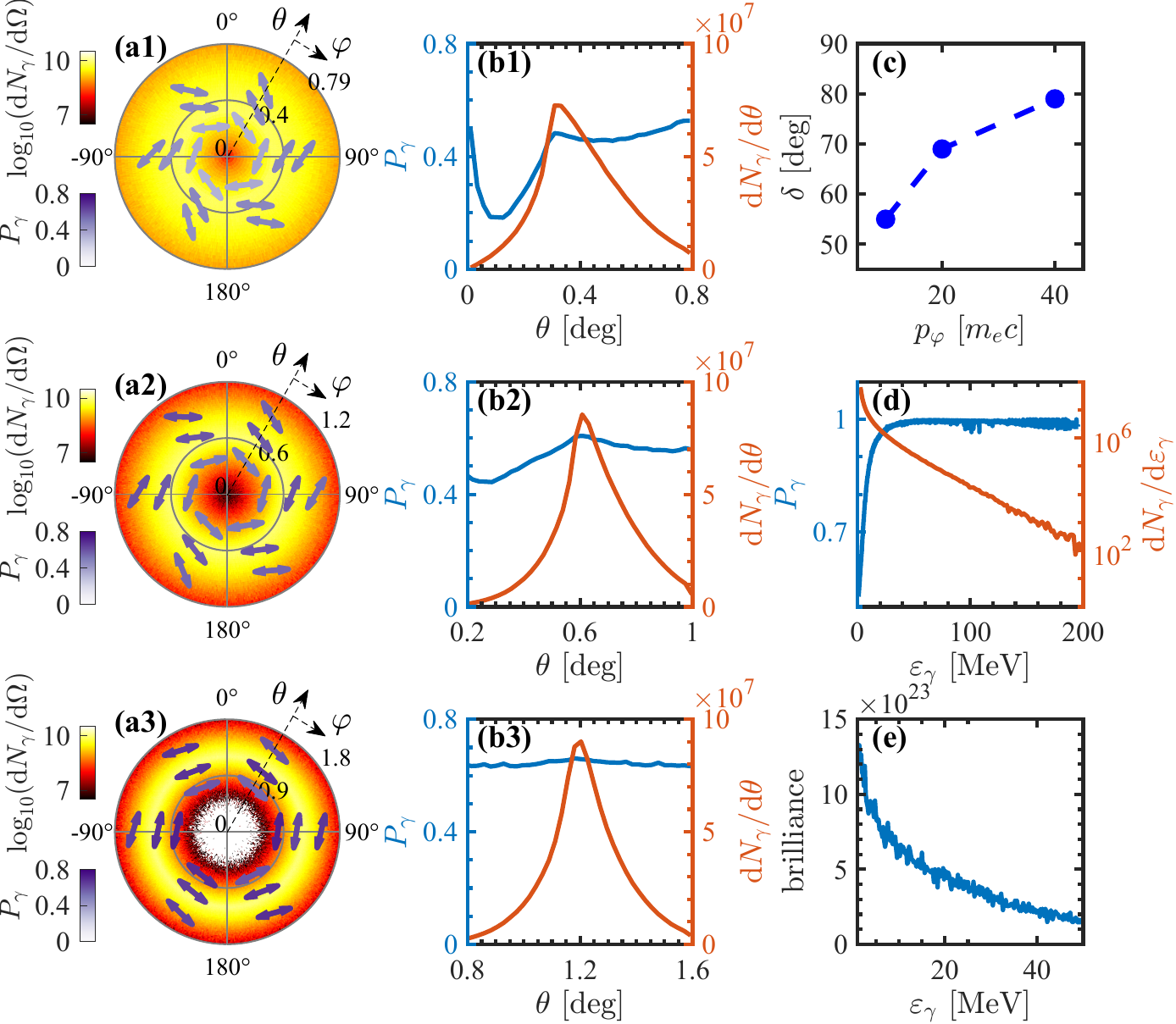}
	\begin{picture}(300,0)	
	\end{picture}
	\caption{  (a1)-(a3) Angle resolved distribution $\log_{10}\left({\rm d} N_\gamma/{\rm d}{\rm  \Omega} \right)$ (background heatmap) and average polarization $P_\gamma$ of the emitted $\gamma$ photons with respect to the polar angle $\theta$ and the azimuth angle $\varphi$. Here, ${\rm d}{\rm \Omega} = \sin \theta {\rm d} \theta {\rm d} \varphi$, $\theta$ is the angle between the photon momentum and the $+x$-axis, and $\varphi$ is the angle between the projection of the momentum onto the $yz$-plane and the $+y$-axis. The superimposed double-headed arrows indicate the average polarization direction, while their color represents the degree of polarization $P_\gamma$. 
(b1)-(b3) Angle-resolved polarization degree $P_\gamma$ (blue) and distribution ${\rm d} N_\gamma/{\rm d}\theta$ (red) of all emitted $\gamma$ photons vs $\theta$. 
Here, panels (a1, b1), (a2, b2), (a3, b3) correspond to the case with an initial electron azimuthal momentum of $p_{\varphi}=10m_ec$, $20m_ec$, $40m_ec$, respectively. (c) The angle $\delta$ as a function of the electron initial azimuthal momentum $p_{\varphi}$. (d) Energy-resolved polarization degree $P_\gamma$ (blue) and distribution ${\rm d}N_\gamma/{\rm d}\varepsilon_\gamma$ (red) of $\gamma$ photons within $0.15^\circ<\theta<0.65^\circ$ vs the photon energy $\varepsilon_\gamma$ for $p_{\varphi}=20m_ec$. (e) The brilliance [photons$/(\mathrm{s\,mm^2mrad^2\times0.1\%bandwidth})$] of the $\gamma$ rays as a function of the the photon energy $\varepsilon_\gamma$. 
}
	\label{fig2}
\end{figure*}

The simulation box has dimensions of $x \times y \times z = 7\, \mu\mathrm{m} \times 10 \,\mu\mathrm{m} \times 10 \,\mu\mathrm{m}$, with grid cells of $560\times400 \times400$ cells. A rotating electron beam with an energy of 1~GeV, a charge of 1.73 nC, a relative energy spread of 5\%, and an angle spread of $0.5^\circ$, propagates along the +$x$ axis. Given that an electron beam with azimuthal momentum possesses a hollow density profile as a result of centrifugal forces, we specify the initial beam density in our simulations as $n_{\rm b} = n_{\rm {b0}}\exp\left[-(x-x_{\rm c})^2/\sigma_x^2 - (r - r_{\rm c})^2/\sigma_{\perp}^2\right]$. Here, $r=\sqrt{y^2+z^2}$, $\sigma_x=0.5\ \mu\mathrm{m}$, $\sigma_{\perp}=1\ \mu\mathrm{m}$, and $n_{\rm{b0}}$ is the maximum density of the beam. The electron beams with comparable parameters can be produced using advanced conventional accelerators or laser-driven accelerators~\cite{babjak2024direct, aniculaesei2023acceleration, Clarke2022advanced, yakimenko2019FACET, yakimenko2019prospect}, as detailed in the Supplementary Material~\cite{SM}. 
A solid aluminum foil with density $n_{\rm{Al}}=5.4\times10^{19}\ \mathrm{cm}^{-3}$ and thickness $d=0.5\,\mu\mathrm{m}$ is employed with its front surface at $x=5\,\mu\mathrm{m}$. In this scheme, thin, low-Z targets are preferred to suppress unwanted energy loss via bremsstrahlung and Bethe-Heitler processes, since both cross-sections scale approximately as
$Z^2$~\cite{heitler1984quantum}. A moving window is used, starting at $t=8.25 \ \rm{fs}$ and moving along the $+x$-axis at a speed of $c$. In the simulations, each cell contains 10 macroparticles for beam electrons and aluminum atoms. Additionally, field ionization is modeled using a hybrid approach that incorporates both tunnel ionization~\cite{Ammosov1986tunnel} and barrier-suppression ionization~\cite{posthumus1997molecular}.

\section{Generation and Properties of Hybrid CV $\gamma$ Rays}
The main results for the generated $\gamma$ rays are summarized in Figure~\ref{fig2}, which compares three cases with different initial azimuthal momenta of the electron: $p_{\varphi} = 10m_{\rm e}c$ (a1, b1), $20m_{\rm e}c$ (a2, b2), and $40m_{\rm e}c$ (a3, b3). As shown in Figure~\ref{fig2}(a), the emitted $\gamma$ rays exhibit hybrid CV polarization. With increasing $|p_{\varphi}|$, the radial polarization component gradually decreases, whereas the azimuthal polarization component gradually increases. Since the angle-resolved distribution and polarization degree of the emitted $\gamma$ rays exhibit cylindrical symmetry, we analyze their dependence on the polar angle $\theta$, as shown in Figure~\ref{fig2}(b). As $|p_{\varphi}|$ increases, the polar angle $\theta$ at the peak of $\gamma$-ray number rises approximately linearly with $|p_{\varphi}|$, reaching $0.3^\circ$, $0.6^\circ$, and $1.2^\circ$ for $p_{\varphi} = 10m_{\rm e}c$, $20m_{\rm e}c$, and $40m_{\rm e}c$, respectively. In all cases, most of the $\gamma$ rays distribute within a $\theta$ width of $0.8^\circ$. Furthermore, the degree of polarization increases with $|p_{\varphi}|$, with average polarization degrees of 0.44, 0.57, and 0.65 for the cases shown in Figures~\ref{fig2}(b1), (b2), and (b3), respectively. This enhancement occurs because a larger $p_{\varphi}$ more effectively suppresses the polarization cancellation, which will be discussed in detail in Figure~\ref{fig3}. The result of average polarization degrees is comparable to that achieved in laser-electron collision schemes ($\sim$60\%)\cite{li2020polarized, Xue2020Generation}.

The hybrid CV polarization state is characterized by a polarization angle $\delta$, defined as the angle between the polarization direction $\mathbf{\hat e}_{P}$ and the radial unit vector $\mathbf{\hat e}_{r}$. Here, $\delta=0^\circ$ corresponds to purely radial polarization, while $|\delta|=90^\circ$ corresponds to purely azimuthal polarization. Our simulations show that $|\delta|$ increases with $|p_{\varphi}|$, reaching $55^\circ$, $69^\circ$, and $79^\circ$ at $p_{\varphi} = 10m_{\rm e}c$, $20m_{\rm e}c$, and $40m_{\rm e}c$, respectively [Figure~\ref{fig2}(c)], with a limit of $90^\circ$. This dependence makes it possible to diagnose the azimuthal momentum of short-pulse electron beams, which remains a great challenge in current laboratories~\cite{thaury2013angular}. Since the sign of $\delta$ is determined by the sign of $p_{\varphi}$, the angle $\delta$ can cover the range $(-90^\circ,\,90^\circ)$. 

We now discuss the energy dependence of the $\gamma$-ray properties, using the case of $p_{\varphi} = 20m_{\rm e}c$ as an example. Figure~\ref{fig2}(d) shows the energy-resolved polarization $P_\gamma$ and energy spectrum ${\rm d}N_\gamma/{\rm d}\varepsilon_\gamma$ of photons within $0.15^\circ<\theta<0.65^\circ$ for the case of $p_{\varphi}=20m_ec$. These $\gamma$ rays exhibit an exponential energy spectrum with a cutoff at approximately 200 MeV. The average polarization increases with photon energy $\varepsilon_\gamma$ from $\sim 60\%$ at the MeV range to nearly 100\% above 25 MeV. This trend is consistent with the theoretical prediction, as shown in Figure~\ref{fig3}(c), that $P_\gamma$ increases with $\varepsilon_\gamma / \varepsilon_e$ over our parameter range ($\varepsilon_{\gamma}/\varepsilon_{e} < 0.2,\ \varepsilon_{e} \sim 1 \  \rm {GeV}$).
These photons have a root-mean-square angular divergence of 0.79 $\rm mrad$ $\times$ 0.79 $\rm mrad$, and spatial dimensions of $0.22$ $\rm \mu m$ (longitudinal) and $1.20\, {\rm \ \mu m} \times 1.20\, {\rm \mu m}$ (transverse). The corresponding brilliances at $\varepsilon_{\gamma}=1$ MeV, 10 MeV, and 50 MeV are $1.37 \times 10^{24}$, $7.86 \times 10^{23}$, and $1.43 \times 10^{23}$ photons$/(\mathrm{s\,mm^2mrad^2\times0.1\%bandwidth})$, respectively, as shown in Figure~\ref{fig2}(e).

\section{Mechanism of $\gamma$-ray Polarization}
\label{section3}

\begin{figure}	[t]	 
	\setlength{\abovecaptionskip}{-0.2cm}
	\setlength{\belowcaptionskip}{-0.3cm}
	\centering
	\includegraphics[width=1.0\linewidth]{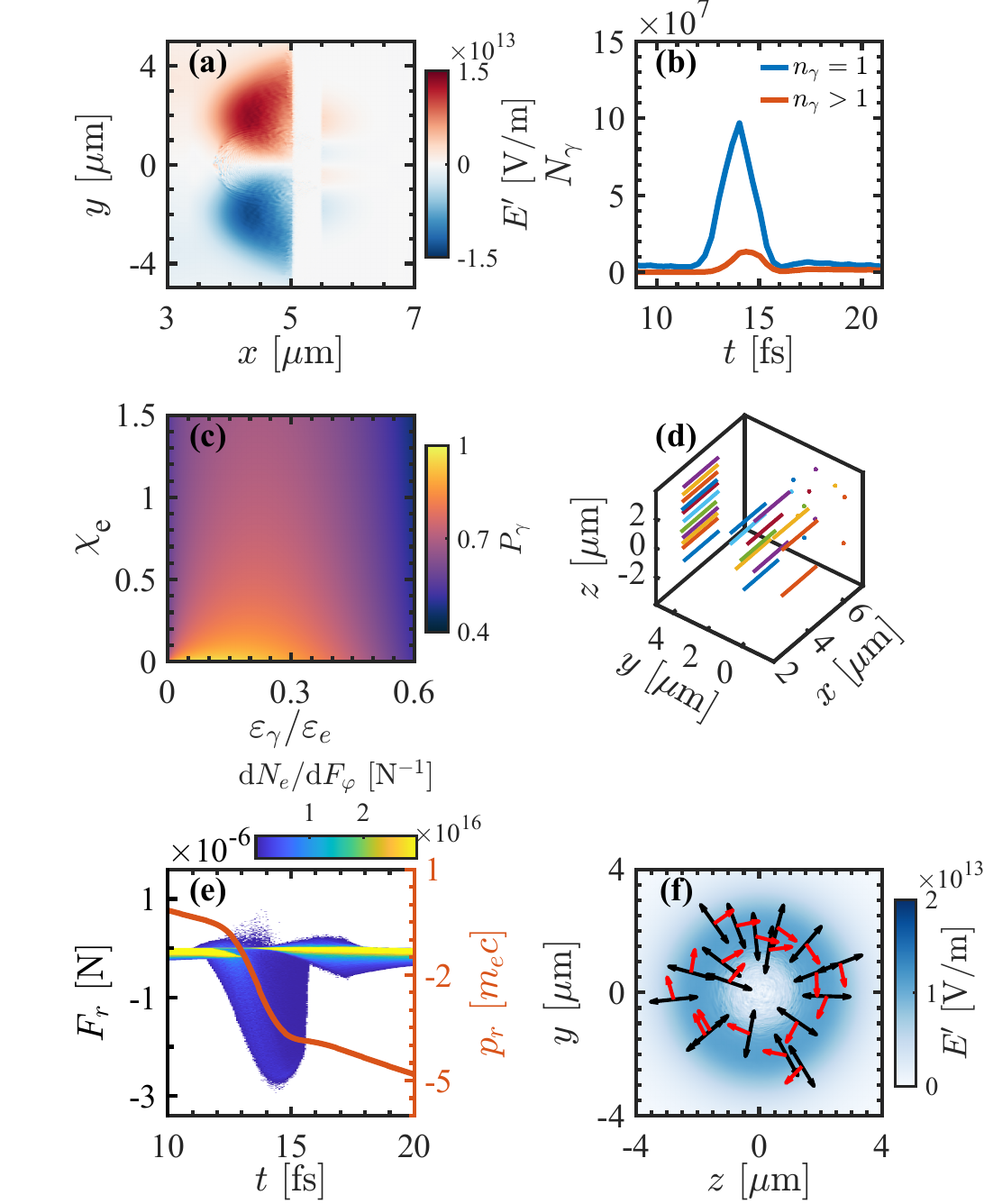}
	\begin{picture}(300,0)

	\end{picture}
	\caption{
		(a) Distribution of the effective electric field $E^{\prime}$ in the xy plane at z=0. (b) Time-dependent photon generation rate, where the blue line represents the rate for first-photon emission and the red line represents the rate for multiple-photon emission. (c) Average polarization degree $P_{\gamma}$ as a function of energy ratio $\varepsilon_{\gamma}/\varepsilon_{e}$ and  QED parameter $\chi_e$. (d) Electron trajectories during interaction with the CTR field and their projections in the $yz$ and $xz$ planes. (e) Distribution of radial forces $F_r$ acting on electrons at different times, with colour representing the number of electrons. The red solid line shows the evolution of the average radial momentum $p_r$ of the beam. (f) Distribution of photon polarization directions (black double arrows) and momentum directions (red arrows) in the $yz$ plane, with the color scale representing the effective electric field $E^{\prime}$.
	}
	\label{fig3}
\end{figure}

When a relativistic electron beam traverses a target, the abrupt change in dielectric constant generates an intense CTR field on both sides of the target~\cite{carron2000fields, han2013vectorial}, as shown in Figure~\ref{fig3}(a). This leads to substantial $\gamma$-ray emission from the beam electrons via the nonlinear Compton scattering process. In a cylindrical coordinate system aligned with the $x$-axis ($\hat{\mathbf{e}}_x$, $\hat{\mathbf{e}}_r$, $\hat{\mathbf{e}}_\vartheta$), this field takes the form $\mathbf{E}_\mathrm{CTR} = E_r\hat{\mathbf{e}}_r + E_x\hat{\mathbf{e}}_x$ and $\mathbf{B}_\mathrm{CTR} = -B_\vartheta\hat{\mathbf{e}}_\vartheta$\cite{sampath2021extremely}. For relativistic electrons propagating along the $x$-axis, the effective electric field can be written as $\mathbf{E}^{\prime} = \mathbf{E}_\perp+\mathbf{v}\times\mathbf{B}\approx(E_r + c B_\vartheta) \hat{\mathbf{e}}_r$, where $\mathbf{E}_\perp$ is the transverse electric field. The intensity of the $\gamma$-ray emission is governed by the nonlinear QED parameter $\chi_e \approx \gamma_\mathrm{e}|\mathbf{E}^{\prime}|/{E}_\mathrm{c}$~\cite{ritus1985quantum}, where $E_{c}=m_{e}^{2}c^{3}/(|e|\hbar)\approx1.3\times10^{18}\:\mathrm{V/m}$ is the Schwinger critical field. The CTR field comprises components propagating along both the $-x$ and $+x$ directions. Electrons interacting with the field propagating along $-x$ experience a large $\chi_e$, thereby dominating the $\gamma$-ray emission, while the contribution from the field propagating along $+x$ is negligible.

As depicted in Figure~\ref{fig3}(b), high-energy photons primarily originate from single photon emissions from electrons, with multiple photon emissions accounting for only about 5.5\% of the total photons. This is due to the significant reduction in electron energy and the corresponding decrease in $\chi_e$ after the first emission. Consequently, we consider only the first photon emitted by an electron. For an initially unpolarized electron beam, the initial spin averages to zero ($\overline{\mathbf{S}}_i = 0$). This leads to a simplification of Eqs.~\eqref{eq:xi_bar}, yielding $\overline{\xi_1} = \overline{\xi_2} = 0$ and a non-zero $\overline{\xi_3} = K_{\frac{2}{3}}(\rho)/w$. This result indicates that the emitted photon is linearly polarized along the axis $\hat{\mathbf{e}}_1$ of the instantaneous frame. In this case, the degree of polarization is given by $P_\gamma = \overline{\xi_3} = (1-u){K}_{2/3}(\rho)/[- (1-u){{\rm Int} K}_{1/3}(\rho) + (u^2-2u+2){ K}_{2/3}(\rho)]$. As illustrated in Figure~\ref{fig3}(c), $P_\gamma$ decreases with increasing $\chi_e$ at a fixed $\varepsilon_{\gamma}$; conversely, at a fixed $\chi_e$, $P_\gamma$ first increases with the photon energy $\varepsilon_{\gamma}$, reaches a maximum, and then decreases to zero. In our scheme, since the emitted $\gamma$ photons have small polar angles and the electron's trajectory during the interaction is approximately linear [Figure~\ref{fig3}(d)], we can approximate the velocity of the parent electron as $\hat{\mathbf{v}}\approx \hat{\mathbf{e}}_x$ and its transverse acceleration as $\hat{\mathbf a}_\perp \approx \mathbf{E}'/|\mathbf{E}'| \approx -\hat{\mathbf{e}}_r$. The $\hat{\mathbf{e}}_1$ axis of the instantaneous frame becomes $\hat{{\mathbf{e}}}^0_1 =\hat{\mathbf a}-\hat{\mathbf v}(\hat{\mathbf v}\cdot\hat{\mathbf a})  \approx -\hat{{\mathbf e}}_r \parallel -\mathbf{E}'$, which shows that the emitted $\gamma$ photon is polarized along the radial direction. The remaining axes of the instantaneous frame are $\hat{\mathbf{k}}_\gamma\approx\hat{\mathbf v}$ and $\hat{{\mathbf{e}}}^0_2 = \hat{\mathbf v} \times \hat{\mathbf a} \approx -\hat{{\mathbf{e}}}_\vartheta$. As all photons with a given wavevector $\hat{\mathbf{k}}_\gamma$ share this common instantaneous frame ($\hat{\mathbf{k}}_\gamma$, $\hat{\mathbf{e}}_1^0$, $\hat{\mathbf{e}}_2^0$),  we adopt it as our observation frame ($\hat{\mathbf{k}}_\gamma$, $\hat{\mathbf{o}}_1$, $\hat{\mathbf{o}}_2$).

The hybrid $\gamma$-ray polarization originates from the misalignment between their transverse momentum and polarization directions. During the interaction with the CTR field, an electron experiences a radial Lorentz force $\mathbf{F}_{r}$, and acquires a radial momentum component [$\sim 3 m_ec$ in Figure~\ref{fig3}(e)]. The azimuthal and radial momentum components of the parent electron ($p_\varphi$ and $p_r$) are subsequently transferred to the emitted photon. Therefore, the direction of the photon transverse wave vector ${\mathbf k}_\perp$ deviates from the radial direction by an angle $\arctan(p_\varphi / p_r)$ [Figure~\ref{fig3}(f)]. As a result, for a given observation angle $\phi$, the detected photons originate from electrons at a specific position $\vartheta = \phi - \arctan(p_\varphi / p_r)$, and their observed polarization is oriented at an angle $-\arctan(p_\varphi / p_r)$ relative to $\hat{\mathbf{e}}_r(\phi)$. Consequently, the polarization angle $\delta$ can be approximated as $\delta \approx \arctan(p_\varphi / p_r)$. By varying the initial azimuthal momentum $p_\varphi$, $\delta$ can be continuously tuned, yielding the different polarization states shown in Figure~\ref{fig2}(a). Furthermore, since the intensity of the CTR field is proportional to the density of the driving electron beam\cite{sampath2021extremely}, adjusting the beam density changes $\mathbf{F}_{r}$ and $p_r$, thereby providing an additional choice for manipulating the polarization states.

\section{Parameter Influence}

\begin{figure}	[t]	 
	\setlength{\abovecaptionskip}{-0.2cm}
	\setlength{\belowcaptionskip}{-0.3cm}
	\centering
	\includegraphics[width=1.0\linewidth]{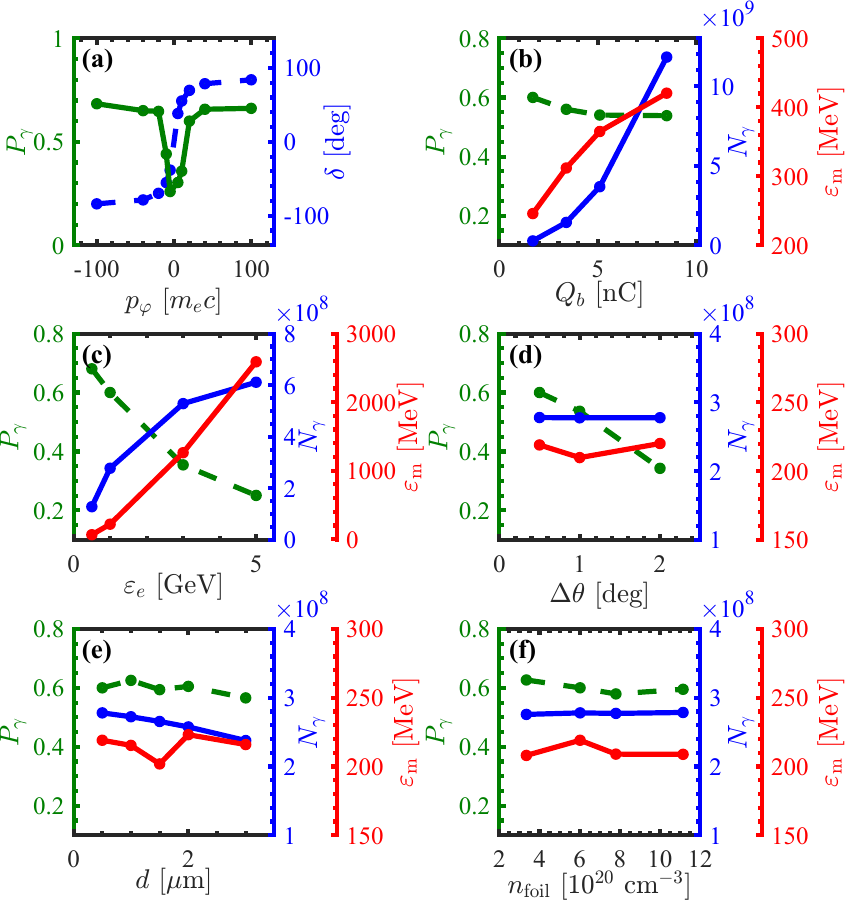}
	\begin{picture}(300,0)

		\end{picture}
	\caption{
(a) Average polarization degree $P_{\gamma}$ (green line) and polarization angle $\delta$ (blue line) as a function of azimuthal momentum $p_\varphi$. Effects of (b) charge of the electron beam $Q_b$, (c) energy of the electron beam $\varepsilon_e$, (d) beam angle spread $\Delta\theta$, (e) thickness of the foil, and (f) density of the foil on average polarization degree $P_{\gamma}$ (green line), number of the emitted photons $N_\gamma$ (blue line), and cutoff energy $\varepsilon_{\rm{m}}$ (red line).
	}
\label{fig4}
\end{figure}

To demonstrate experimental feasibility, the impact of key parameters of the electron beam and the foil on the $\gamma$-ray polarization degree $P_{\gamma}$, number of the emitted photons $N_\gamma$, and cutoff energy $\varepsilon_{\rm{m}}$ is summarized in Figure~\ref{fig4}. Here, only one parameter is changed in each case, while all other parameters are kept fixed. As the electron azimuthal momentum $p_\varphi$ increases, the polarization angle $\delta$ increases from approximately -90$^\circ$ to 90$^\circ$[Figure~\ref{fig4}(a)]. This arises from the increase in $\arctan(p_\varphi/p_r)$, as the radial momentum $p_r$ remains nearly constant. The scaling relationship between $p_\varphi$ and $\delta$ provides a potential method for diagnosing electron azimuthal momentum through $\gamma$-ray polarization state in laboratory. The average polarization degree $P_{\gamma}$ rises with $|p_\varphi|$ and reaches 70\%, since a larger $|p_\varphi|$ suppresses the polarization cancellation~\cite{cao2025generating}. A higher electron charge $Q_b$ strengthens the CTR field inntensity, leading to an enhancement of $\mathbf{E}^{\prime} \propto Q_b$. This subsequently raises $\chi_e \approx \gamma_\mathrm{e}|\mathbf{E}^{\prime}|/{E}_\mathrm{c}$ and the radiation probability\cite{li2020polarized, Xue2020Generation}. As shown in Figure~\ref{fig4}(b), increasing $Q_b$ leads to an increase in both the cutoff energy $\varepsilon_{\rm{m}} \propto \chi_e \gamma_e$ and the number of radiated photons $N_\gamma$, while $P_\gamma$ decreases (as discussed in Sec.~\ref{section3}). Similarly, increasing electron beam energy $\varepsilon_e$ also raises $\chi_e$, leading to consequently increasing $N_\gamma$ and $\varepsilon_{\rm{m}}$ while decreasing $P_{\gamma}$, as illustrated in Figure~\ref{fig4}(c). Conversely, an increase in the angle spread $\Delta\theta$ broadens the electron transverse momentum distribution, enhancing the polarization cancellation~\cite{cao2025generating}. This leads to a reduction in both the polarization degree, whereas the cutoff energy $\varepsilon_{\rm{m}}$ and the total photon yield $N_\gamma$ remain stable, as shown in Figure~\ref{fig4}(d).

The influence of foil parameters is also investigated. As the foil thickness $d$ increases, the number of emitted photons $N_\gamma$ decreases slightly, while both the polarization degree $P_{\gamma}$ and cutoff energy $\varepsilon_{\rm{m}}$ remain almost unchanged, as shown in Figure~\ref{fig4}(e). They spend more time traversing a thicker foil, thereby reducing the time they interact with the CTR field propagating along the $+x$ direction and resulting in fewer $\gamma$-ray emission by the same time. In contrast, the density of the target has a negligible impact on $N_\gamma$, $P_{\gamma}$, and $\varepsilon_{\rm{m}}$, according to Figure~\ref{fig4}(f), since it has little influence on the CTR field~\cite{sampath2021extremely}.

\begin{figure}	[t]	 
	\setlength{\abovecaptionskip}{-0.2cm}
	\setlength{\belowcaptionskip}{-0.3cm}
	\centering
	\includegraphics[width=1.0\linewidth]{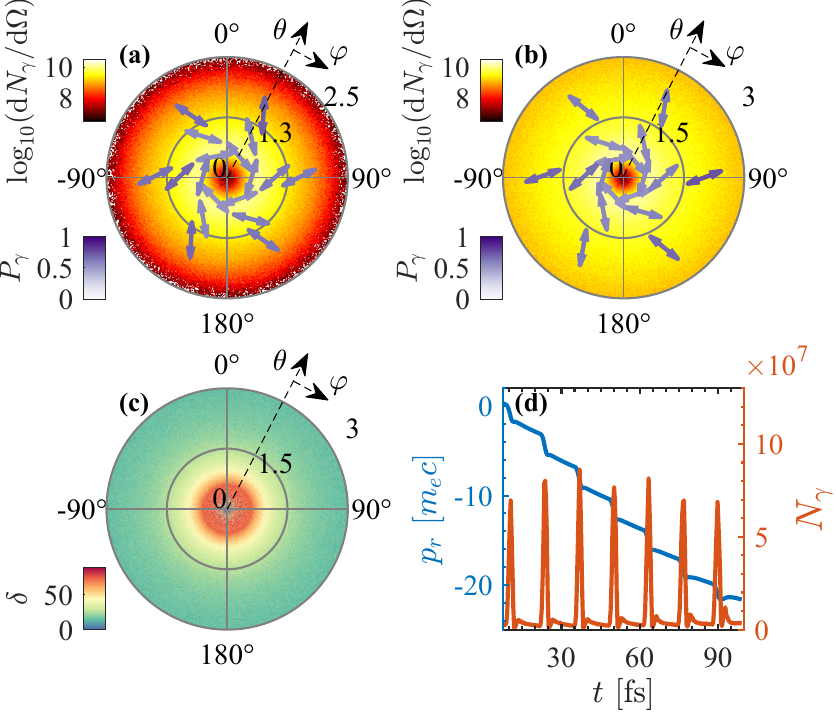}
	\begin{picture}(300,0)

		\end{picture}
	\caption{
	(a)-(b) Angle-resolved distribution $\log_{10}\left({\rm d} N_\gamma/{\rm d}{\rm  \Omega} \right)$ (background heatmap) and average polarization $P_\gamma$ of the emitted $\gamma$ rays, where (a) corresponds to electrons traversing 3 foils and (b) 7 foils. (c) Angle-resolved distribution of $\gamma$-ray polarization angle $\delta$, after traversing 7 foils. (d) Evolution of the average radial momentum of electrons (blue line) and the number of radiated photons (red line).
}
\label{fig5}
\end{figure}

We further investigate the interaction of the rotating electron beam with multiple foils. The results reveal a more complex polarization structure that can be controlled efficiently by varying the number of foils. For example, we use the electron beam with $p_\varphi = 20m_ec$ and the foil separation of 4 $\rm {\mu m}$. Here, all other parameters are identical to Sec. 3. As shown in Figures~\ref{fig5}(a) and (b), the radial polarization component of the emitted $\gamma$ rays increases with the polar angle $\theta$ for a given azimuthal angle $\varphi$. This $\theta$-dependent polarization structure becomes clearer with increasing the foil number, owing to the broader angular distribution of the emitted $\gamma$ rays. For both $N_{\text{foil}} = 3$ [Figure~\ref{fig5}(a)] and $N_{\text{foil}} = 7$ [Figure~\ref{fig5}(b)], the polarization degree $P_\gamma$ remains above 50\%. Figure~\ref{fig5}(c) displays the angular distribution of polarization angle of $\gamma$ rays for $N_{\text{foil}} = 7$. The results reveal a clear decrease of $\delta$ from nearly $90^\circ$ to approximately $10^\circ$ with increasing polar angle $\theta$. Specifically, the polarization is predominantly azimuthal at small $\theta$, whereas it changes to predominantly radial at large $\theta$.

Figure~\ref{fig5}(d) reveals the origin of this polarization structure. Upon interacting with each foil, electrons gain radial momentum and emit $\gamma$ rays through the CTR fields generated at each foil. As  $p_r$ accumulates with increasing the foil number, $p_\varphi/p_r$ gradually decreases. At $N_{\text{foil}}=7$, $p_r$ eventually exceeds $p_\varphi$. This leads to a corresponding reduction in the polarization angle $\delta$ and an increase in the polar angle of the emitted $\gamma$-rays. Consequently, $\gamma$ rays generated at lower $N_{\text{foil}}$ exhibit larger $\delta$ and concentrate around $\theta=0$ in the angular distribution, while those generated at higher $N_{\text{foil}}$ have smaller $\delta$ and appear at larger $\theta$. This gives rise to the polarization structure observed in Figures.~\ref{fig5}(a) and (b). The results demonstrate that the polarization is tunable by simply changing the foil number, thereby making it feasible in experiments.

\section{Conclusion}

In conclusion, we have put forward a novel method for generating polarization-tunable hybrid CV $\gamma$ rays through the interaction of rotating electron beams with solid foils. Our approach achieves continuous tuning of the polarization angle $\delta$ across the range of $(-90^\circ, 90^\circ)$ by adjusting the beam azimuthal momentum $p_\varphi$. We establish a connection between $p_\varphi$ and the polarization structure of the emitted $\gamma$ rays, making it a potential probe for detecting the azimuthal momentum of driving electrons. Our work could offer a promising way for applications of structured $\gamma$ rays in diverse areas, such as high-energy physics, nuclear science, and laboratory astrophysics.

\appendix{}

\sloppy{}

\begin{center}\textbf{Acknowledgement}\end{center}

\noindent  This work is supported by the National Natural Science Foundation of China (Grant Nos. 12135009, 12375244, 12175310, 12505235), the Natural Science Foundation of Hunan Province of China (Grant No. 2025JJ30002), the China Postdoctoral Science Foundation (Grant No. 2024M762568), the Postdoctoral Fellowship Program of CPSF (Grant No. GZC20252248), and the Fundamental Research Funds for the Central Universities of Ministry of Education of China (Grant No. xzy012025079).

\vspace*{4mm}

\bibliographystyle{unsrt}
\bibliography{mybib}
\end{document}


\author[]{Si-Man Liu$^{1, }$\thanks{These authors contributed equally to this work.}}
\author[]{Yue Cao$^{1, *}$}
\author[]{Kun Xue\textsuperscript{2, }\thanks{xuekun@xjtu.edu.cn}\ }
\author[1]  {Li-Xiang Hu}
\author[1]  {Xin-Yu Liu}
\author[1]  {Xin-Yan Li}
\author[1]  {Chao-Zhi Li}
\author[1]  {Xin-Rong Xu}
\author[1]  {Ke Liu}
\author[1]  {Wei-Quan Wang}
\author[1]  {De-Bin Zou}
\author[1]  {Yan Yin}
\author[2, 3]  {Jian-Xing Li}
\author[]{Tong-Pu Yu$^{1,}$\thanks{tongpu@nudt.edu.cn}}

\address[1]{College of Science, National University of Defense Technology, Changsha 410073, China}
\address[2]{Ministry of Education Key Laboratory for Nonequilibrium Synthesis and Modulation of Condensed Matter, State Key Laboratory of Electrical Insulation and Power Equipment, Shaanxi Province Key Laboratory of Quantum Information and Quantum Optoelectronic Devices, School of Physics, Xi'an Jiaotong University, Xi'an 710049, China}
\address[3]{Department of Nuclear Physics, China Institute of Atomic Energy, P.O. Box 275(7), Beijing 102413, China}

\newtheorem{theorem}{Theorem}
\shortauthor{S. M. Liu et al.}
\shorttitle{Supplemental Material}
\title{Generation of Polarization-Tunable Hybrid Cylindrical Vector $\gamma$ Rays from Rotating Electron Beams
           Supplemental Material}

\maketitle

\section{Generation and Feasibility Validation of Rotating Electron Beams}

Rotating electron beams, which carry significant azimuthal momentum, can be generated through various laser-driven schemes. A straightforward method involves the interaction of Laguerre-Gaussian (LG) laser pulses with plasma~\cite{Hu2024Rotating, Shi2024Advances}, where electrons gain substantial azimuthal momentum from the helical wavefront structure, resulting in hollow beam profiles. For instance, the work by Hu et al. demonstrated that twisted laser pulses interacting with micro-droplet targets can generate rotating relativistic electron sheets with azimuthal momentum up to tens of $m_e c$. These electron sheets can achieve peak densities exceeding $1.5 \times 10^{22}$ cm$^{-3}$ at a radius of approximately 1 $\mu m$, with cutoff energy reaching 340 MeV. LG laser-driven wakefields can also produce rotating electron beams with hundred-MeV energy, but the azimuthal momentum and charge are substantially lower than the former direct acceleration scheme~\cite{shen2017acceleration, zhang2016acceleration}. Beyond the LG laser approach, multiple Gaussian laser pulses with twisted pointing directions could alternatively generate such beams~\cite{Shi2023Efficient}. By transferring orbital angular momentum from the twisted Gaussian beams to the plasma, electrons could potentially acquire azimuthal momenta on the order of $m_e c$ in strong axial magnetic fields exceeding 10 kT. Additionally, a single Gaussian laser beam interacting with plasma lenses could achieve similar beam-splitting effects~\cite{Zhang2025Plasma}.

\begin{figure}
\centering
\includegraphics[width=1\linewidth]{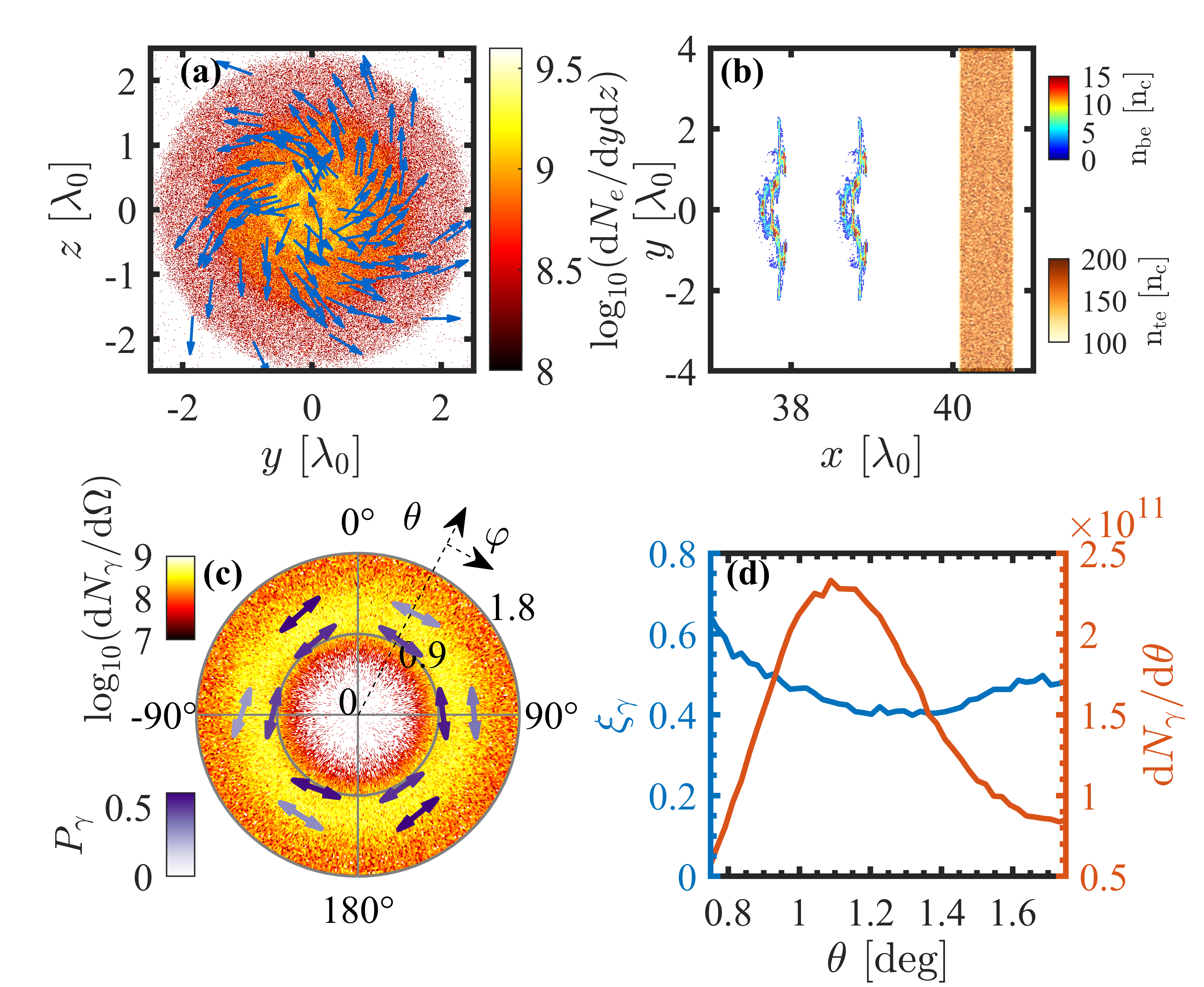}
\caption{(a) Electron sheet distribution $\log_{10}\left({\rm d} N_e/{\rm d}{x}{\rm d}y \right)$ in the yz plane from LG laser-plasma interaction, where blue arrows indicate the transverse momentum directions of electrons. (b) Initial density distribution of electron beams and foil in the xy plane at z=0. (c) Angle resolved distribution $\log_{10}\left({\rm d} N_\gamma/{\rm d}{\rm  \Omega} \right)$ (background heatmap) and average polarization $P_\gamma$ of the emitted $\gamma$ photons with respect to the polar angle $\theta$ and the azimuth angle $\varphi$. Here, ${\rm d}{\rm \Omega} = \sin \theta {\rm d} \theta {\rm d} \varphi$, $\theta$ is the angle between the photon momentum and the $+x$-axis, and $\varphi$ is the angle between the projection of the momentum onto the $yz$-plane and the $+y$-axis. The superimposed double-headed arrows indicate the average polarization direction, while their color represents the degree of polarization $P_\gamma$. (d) Angle-resolved polarization degree $P_\gamma$ (blue) and distribution ${\rm d} N_\gamma/{\rm d}\theta $ (red) of all emitted $\gamma$ photons vs $\theta$. }
\label{figS1}
\end{figure}

To verify the feasibility of our scheme, we take the LG laser-plasma interaction scheme for example and performed spin-resolved three-dimensional PIC simulations using SLIPs. We employ a left-handed circularly polarized LG laser pulse with mode (1, 0), wavelength $\lambda_0 = 1\ \mu m$, dimensionless peak amplitude $a_0 = 45$, carrier-envelope phase (CEP) $\Phi_0 = 3\ \pi/2$ propagates along the x-axis and is focused onto a pre-ionized helium micro-droplet target with a radius of $\lambda_0$ and density of 10$n_c$. Here, $n_c=1.12 \times 10^{21}\ cm^{-3}$ is the critical density for a $\lambda_0 = 1\ \mu m$ laser pulse, and $\Phi_0$ denotes the carrier-envelope phase that controls the timing between the laser pulse envelope and its carrier wave.  An electron sheet exhibiting significant azimuthal momentum is successfully generated, with its density and momentum distributions in the yz-plane shown in Figure~\ref{figS1} (a).

Due to the short pulse duration of the electron sheet, a single sheet cannot sufficiently interact with the coherent transition radiation (CTR) field as it traverses the foil. Using multiple electron sheets can significantly enhance the $\gamma$-ray emission. The leading sheets drive the CTR field, while the following sheets emit $\gamma$ rays through interaction with this CTR field. Such multiple electron sheets can be generated by adjusting the laser pulse duration and CEP. For simplicity, we consider two sheets separated by $1\ \mu$m interacting with the foil to verify the generation of hybrid cylindrical vector (CV) $\gamma$ rays. Here, the foil parameters are identical to those used in the main text. Figure~\ref{figS1} (b) shows the initial density distributions of the electron sheets and foil, where the sheets propagate along the $+x$ direction. The emitted $\gamma$ rays exhibit a hybrid CV polarization with an average $\delta \approx 70$, as shown in Figure~\ref{figS1} (c). This is highly consistent with the $\gamma$-ray structure generated by the electron beam with $p_{\varphi} = 40m_ec$ in Figure 2(a3). Since the angle-resolved distribution $\log_{10}\left({\rm d} N_\gamma/{\rm d}\Omega \right)$ and polarization $P_\gamma$ of the emitted $\gamma$ rays exhibit cylindrical symmetry [Figure~\ref{figS1}(c)], their dependence on the polar angle $\theta$ is analyzed. The polarization degree of the emitted $\gamma$ rays exceeds 40\% for all polar angles, with an average polarization degree of 45\%, as shown in Figure~\ref{figS1}(d). The hybrid CV $\gamma$ rays are successfully generated using rotating electron beams produced by the established LG laser-plasma interaction scheme, thereby confirming the feasibility of our method.

\begin{figure}
\centering
\includegraphics[width=1\linewidth]{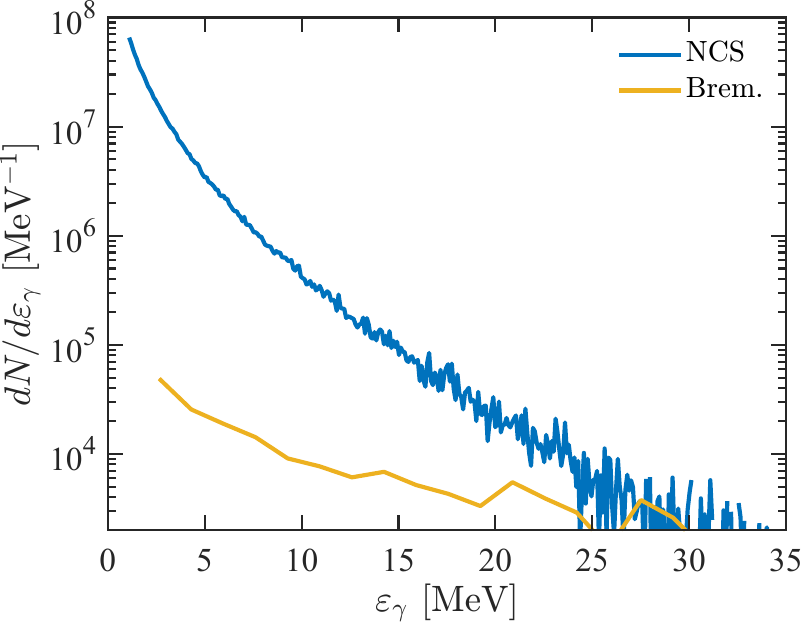}
\caption{Photon energy spectra from nonlinear Compton scattering (NCS) and bremsstrahlung processes, respectively.}
\label{figS2}
\end{figure}

\section{Effects of Bremsstrahlung and Bethe-Heitler Processes}
During the beam target interaction, the Bremsstrahlung and Bethe-Heitler processes may come into play. The cross-section for an electron with kinetic energy $E_{kin}$ to produce a bremsstrahlung photon with energy exceeding $E_0$ is given by $\sigma_{\gamma}=1.1\times10^{-16}Z^{2}\left[0.83\left(\frac{E_{0}}{E_{kin}}-1\right)-\ln\frac{E_{0}}{E_{kin}}\right]cm^{2}$\cite{heitler1984quantum}. Conversely, the total cross-section for the Bethe-Heitler pair production process is $\sigma_{\mathrm{BH}}=\frac{28}{9}Z^{2}\alpha r_{\mathrm{e}}^{2}\left(\ln\frac{2\varepsilon_{\gamma}}{m_{\mathrm{e}}c^{2}}-\frac{109}{42}\right)$ for $\varepsilon_\gamma\gg m_ec^2$\cite{heitler1984quantum}. Here, $Z$ is the atomic number of the target, $\alpha$ is the fine-structure constant, $r_e$ and $m_e$ represent the radius and mass of electron, respectively, $\varepsilon_{\gamma}$ is the photon energy, and $c$ is the speed of light in vacuum. In our scheme, the low-Z target suppresses both processes, as their cross-sections are proportional to $Z^2$. 

To ensure the reliability of our results presented in the main text, we employ the 3D PIC code EPOCH to evaluate the contributions of the Bethe-Heitler and bremsstrahlung processes in the beam-target interaction.  We perform simulations using the parameters corresponding to Figure 2(a2). No Bethe-Heitler process is observed in the simulations, which is attributed to the small cross section associated with the low photon average energy and the low-Z target. Figure ~\ref{figS2} shows the photon energy spectra from NCS and bremsstrahlung processes. It is shown that the bremsstrahlung process contributes only 0.2\% of the total photon number and 5\% of the total photon energy, which can be natrually negligible as presented in the main text.

\sloppy{}

\vspace*{4mm}

\bibliographystyle{unsrt}
\bibliography{mybib_sup}